\documentclass[final,5p,times,twocolumn]{elsarticle}
\usepackage{epsfig}
\newtheorem{remark}{Remark}[section]

\newtheorem{theorem}{Theorem}[section]

\newtheorem{lemma}{Lemma}[section]

\newproof{pf}{Proof}
\newproof{pot}{Proof of Theorem}

\usepackage{amssymb}

\journal{}
\begin{document}

\begin{frontmatter}

\title{\bf Synchronization of  the cardiac pacemaker model with delayed pulse-coupling}

\author{Marat Akhmet} \ead{marat@metu.edu.tr}

\address{Department of Mathematics, Middle East Technical University, 06531 Ankara, Turkey}

\begin{abstract}
We consider the  integrate-and-fire model of  the cardiac pacemaker with   delayed pulsatile coupling. Sufficient conditions of synchronization  are obtained for identical and non-identical  oscillators.
\end{abstract}

\end{frontmatter}

\section{Introduction} 
 
 In paper \cite{p} C. Peskin  proposed a model of cardiac  pacemaker, where signal  of fire arises not from an outside stimuli, but  in the population of cells itself.  Well known conjectures  of self-synchronization were formulated.  Solution of these conjectures for identical oscillators  \cite{p,ms}  stimulated  intensive investigations \cite{bot}-\cite{timme2}.  
 
Delays arise  naturally  in many  biological  models \cite{ murray}. In particular, they  were   considered   in firefly  models  \cite{buck1} as delay  between  stimulus and response,  and in  continuously  coupled  neuronal oscillators \cite{ko}.  Authors of  \cite{epg}  considered  the phenomenon for the  Mirollo and Strogatz analysis, \cite{ms}. Identical  oscillators were investigated. Two  oscillators dynamics is discussed mathematically, and the multi-oscillatory  system by  computer simulations. It was found that  the excitatory  model of two  units  ``can get  only  out-of-phase synchronization since in-phase synchronization proved to  be not  stable.'' In paper  \cite{gerst}  a  model without leakage  was  discussed, that  is, oscillators increase at a constant  rate  between  moments of firing. It  was found that a periodic solution  is reached after a finite time.  Consequently,  research of  integrate-and-fire models, which  admit  delays and fire in unison  is still  on the agenda.

    In paper  \cite{akhmet1}  we have introduced  a  new  method   of  investigation  of  biological oscillators.  The method seems to be universal  to  analyze quite identical integrate-and-fire oscillators.  In the present article we extend  it  to  the Peskin's model  with  delayed interaction. Conditions  are found, which guarantee  synchronization  of the model.  Our system is different  than that  in  \cite{epg}, since we suppose that the pulse-coupling is instantaneous, if  oscillators are close to  each  and are  near   threshold.   In next  our papers, we plan to  consider other models, varying  types of the  delay  involvement, as well as inhibitory  models  such  that  analogues of  results in \cite{epg} and \cite{van} can  be obtained.    The method of analysis of non-identical oscillators is based on results of theory   of  differential equations with discontinuities at non-fixed moments  \cite{akhmet}-\cite{samo}.

 \section{The couple of identical oscillators}  \label{sec2}
 
 Let  us  start  to  analyze   two identical oscillators, which  satisfy, if they  do not fire,   
  the following  differential  equations 
 \begin{eqnarray}
 \label{1}
 && 	x'_i = S  - \gamma x_i,
 \end{eqnarray}
 where $ 0 \le x_i \le 1, i = 1,2.$ It is assumed that $S,\gamma$ are positive  numbers and  $\kappa = \frac{S}{\gamma} > 1.$ 
 
 When $x_j(t) = 1,$ then  the  oscillator fires,   $x_j(t+) = 0.$ The firing changes  value of the another  oscillator, $x_i,$   such that 
\begin{eqnarray}
 \label{2'}
 && x_i(t+) = 0,\, {\rm if}\, x_i(t) \ge  1  - \epsilon,
  \end{eqnarray} 
and  
	 \begin{eqnarray}\label{2}
x_i(t +\tau+)=  x_i(t +\tau) + \epsilon,\, {\rm if}\, x_i(t) < 1  - \epsilon.
\end{eqnarray}

We have  that 
	\[x_i(s) =  x_i(t) {\rm e}^{-\gamma(s-t)} +  \int_{t}^s {\rm e}^{-\gamma(s-u)}Sdu
\]
near $t.$
 
 In what  follows,  assume that     
\begin{eqnarray}
	&& \frac{\kappa -1}{\kappa -1 + \epsilon} < {\rm e}^{-\gamma \tau}.
	\label{4}
\end{eqnarray} 
Then, from  
	\[\|x_i(s)\| \le  \|x_i(t)\| {\rm e}^{-\gamma(s-t)} +  \int_{t}^s {\rm e}^{-\gamma(s-u)}Sdu \le \]\[(1 - \epsilon){\rm e}^{-\gamma\tau} + \kappa(1 -{\rm e}^{-\gamma\tau}),
\]
 and  $x_i(t) < 1 - \epsilon,$  we  obtain  that    $x_i(s) < 1,$  for all $s \in [t,t+\tau].$ 
 In other words oscillator  $x_i$  does  not  achieve  the threshold  within  interval  $[t, t+\tau],$ if the distance of   $x_i(t)$   to threshold  is more than  $\epsilon.$   This  is  important for the  construction of the prototype map, and makes a  sense of  condition  (\ref{2}).  
 
One must  emphasis   that    couplings   of  units are  not  only  delayed in our model. By (\ref{2'}) oscillators interact  instantaneously, if they  are near  threshold. This assumption is natural as   firing   provokes  another oscillator, which  being  close to threshold ``is ready''   to  react  instantaneously. Otherwise,  the interaction is retarded.   

  Next, we shall construct  the prototype map.  Fix a moment $t = \zeta,$ when  $x_1$ fires, and  suppose  that  oscillators are not synchronized.   In interval $[\zeta, \zeta + \tau]$ oscillator $x_2$ moves by   law   
  $x_2(t) =  x_2(\zeta) {\rm e}^{-\gamma(t-\zeta)} +  \int_{\zeta}^t {\rm e}^{-\gamma(t-u)}Sdu,$ 
and 
\begin{eqnarray}
	&& x_2(\zeta + \tau) =  [x_2(\zeta) - \kappa]  {\rm e}^{-\gamma \tau} +  \kappa.
	\label{4'}
\end{eqnarray}
Denote $t = \eta,$ the firing moment of $x_2,$ then 
\[x_2(\eta) =  
	  [x_2(\zeta + \tau) + \epsilon]  {\rm e}^{-\gamma (\eta - \zeta -  \tau)} +   \kappa[1-   {\rm e}^{-\gamma (\eta - \zeta -  \tau)}].\]
		The equation $x_2(\eta) =1$ implies that   
\begin{eqnarray}
	&& {\rm e}^{-\gamma (\eta - \zeta)} =  \frac{1- \kappa}{x_2(\zeta) - \kappa +\epsilon_1},
	\label{7}
\end{eqnarray}  
where    $\epsilon_1 = \epsilon {\rm e}^{\gamma  \tau }.$
Since  $x_1(\eta) = \kappa[1 - {\rm e}^{-\gamma (\eta - \zeta)}],$ we  have that    
\begin{eqnarray}
	&& x_1(\eta) =   \kappa \frac{1- (x_2(\zeta) +\epsilon_1)}{\kappa -  (x_2(\zeta) +\epsilon_1)}.	
	\label{8}
\end{eqnarray}
Introduce  the  following map 
\begin{eqnarray}
	&& L_D(v,\epsilon) = \kappa \frac{1 - (v +\epsilon_1)}{\kappa -  (v +\epsilon_1)},	
	\label{9}
\end{eqnarray}
such  that  $x_1(\eta)  = L_D(x_2(\zeta)),\epsilon).$ 
 If $t = \xi$ is the next   to  $\eta$ firing moment of  $x_2,$ then one can similarly  find that  
 $x_2(\xi) = L_D(x_1(\eta),\epsilon).$ One can  see that  the map  $L_D$ can  be  useful for  our investigation, since  it  evaluates alternatively  the sequence of values $x_1$ and $x_2$ at firing moments. 

 Take $\tau >0$ so  small that   
\begin{eqnarray}
	&&  {\rm e}^{-\gamma  \tau } > \epsilon.	
	\label{10}
\end{eqnarray}
 From (\ref{10}) it  implies that  $\epsilon_1 < 1.$

 One can evaluate that 
	\[L_D(1- \epsilon_1,\epsilon)  = 0.
\]
 and the derivatives of the map   in  $(0,1 - \epsilon_1)$  satisfy
 
\begin{eqnarray}
	&& L_D'(v,\epsilon) = \kappa \frac{1- \kappa}{(\kappa -  (v +\epsilon_1))^2} <0,  
	\label{12}
	\end{eqnarray}
and 
\begin{eqnarray}
	&&  L_D''(v,\epsilon) =2\kappa \frac{1- \kappa}{(\kappa -  (v +\epsilon))^3} < 0  
	\label{13}
	\end{eqnarray}
We can easily  find that  there is a fixed point of the map,  
  \begin{eqnarray}
	&&  v^*    = (\kappa  - \frac{\epsilon_1}{2}) -  \sqrt{\kappa^2 - \kappa + \frac{\epsilon_1^2}{4}},
	\label{13'}
	\end{eqnarray}
and 	
\begin{eqnarray}
	&&  L_D'(v^*,\epsilon)  < -1.  
	\label{14'}
	\end{eqnarray}
	
	That  is, fixed point  $v^*$ is a repellor. 	
	
\begin{figure}[hpbt]
  \centering  
  \epsfig {file=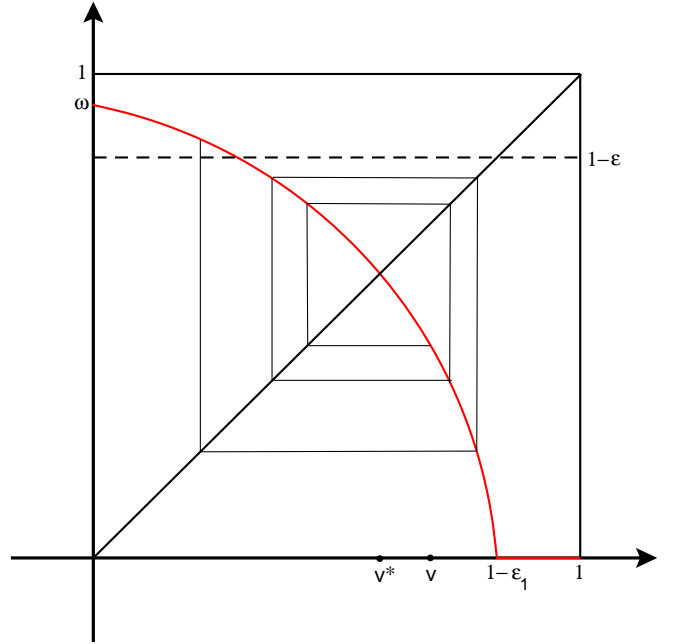, width=3.4in}
  \caption{The graph  of  map  $L_D$ in red, fixed point $v^*,$  and stabilized  trajectory are seen.} \label{Fig1}
\end{figure}

Now, we will define an  extension of $L_D$ on $[0,1]$ in the following way. 
Let 
\begin{eqnarray}
	&&  \omega =  \kappa \frac{1 - \epsilon_1}{\kappa -  \epsilon_1}.	
	\label{11}
\end{eqnarray}
One can see that   $ 1 - \epsilon < \omega < 1,$ if 
\begin{eqnarray}
	&&  {\rm e}^{\gamma  \tau } < \frac{\kappa}{\kappa-1 + \epsilon}.  
		\label{123}
	\end{eqnarray}
In what  follows, we assume that  $\epsilon$ is sufficiently  small such that  (\ref{4}) implies (\ref{123}). We set $L_D(0,\epsilon) = \omega,$ and define $L_D(v,\epsilon) = 0,$ if $1-\epsilon_1 \le v \le 1.$  Since $L_D:[0,1] \to [0,1]$   is  a  monotonic continuous function  and $[0,1]$ is an invariant set of this map, it is   convenient  for analysis  by  using  iterations.  The graph of this map is seen in 	Figure  \ref{Fig1}.	
	To  emphasize   a significance of  this map  for the present analysis, let us see  how iterations of  it  can help to  observe the synchronization. 	
	 Fix $t_0\ge 0,$  a firing moment, $x_1(t_0) = 1, x_1(t_0+) = 0.$ 
	While the couple $x_1,x_2,$ does not synchronize,  there exists a sequence of  moments $t_0 <t_1<\ldots$
	such  that  $x_1$ fires at  $t_i$ with even $i$ and $x_2$ with  odd indices.  Denote $u_i  = x_1(t_i),$ if $i$ is odd, and  $u_i  = x_2(t_i),$ if $i$ is even. One can  easily see that   $u_{i+1} = L_D(u_i,\epsilon), i \ge 0.$  
	The pair synchronizes if and only if there  exists $j \ge 1$ such that $x_1(t) \not = x_2(t),$ if  $t \le t_j,$ and $x_1(t) = x_2(t),$ for $t > t_j.$  In particular,  both oscillators  have to fire at $t_j.$ That is,  inequalities $1- \epsilon \le u_{j-1} < 1$ are valid.   It is  possible if $0 \le u_{j-2} \le L_D^{-1}(1- \epsilon).$ In particular, we have  that   $L_D(0) =  \omega$ satisfies this condition.  In the same time,  if $1- \epsilon_1 \le u_{j-3} \le 1,$ then  $u_{j-2} = 0 = L_D(u_{j-3})$ and  $1- \epsilon < u_{j-1} = \omega < 1$ again.  That  is,  we have found that if there exists an integer $k \ge 0$ such  that $1 -\epsilon \le L_D^k(v) \le 1,$   then the motion $(x_1(t),x_2(t))$  with $x_1(t_0+) =v, x_2(t_0+) =0,$  synchronizes at the $k-$th firing moment.   Conversely,  if a motion $(x_1(t),x_2(t))$  synchronizes, then one can find a firing moment,  $t_0,$ such that  $x_1(t_0+) = 0, x_2(t_0+) =v, v \in [0,1],$ and a number $k$ such that   $1 -\epsilon \le L_D^k(v) \le 1.$
 
  Thus, the last  discussion confirms that  the analysis of synchronization  is consistent fully with the dynamics of the introduced map $L_D(v,\epsilon)$ on $[0,1],$  and  the map  $L_D$  can be applied as the main instrument of the paper. That  is why, we  use  this function  as a prototype  map  in  our investigations.

 	Now, by applying properties of $L_D,$ and analyzing  self-compositions of the map,  one can easily obtain that
  for all  $k \ge 0$  functions $L_D^k$ have only one fixed point, $v^*,$ and  $|[L_D^k(v^*,\epsilon)]'| > 1.$ We skip the discussion as it  is respectively simple, and request a large place. 
  Since all the maps $L_D^k$ have one and the same fixed point, $v^*,$  there is not a $k-$periodic  motion, $k >1,$ of the map. Consequently, for arbitrary point $v \not = v^*$ one has  a stabilized trajectory  present in Figure \ref{Fig1}. The couple synchronizes  when $L_D^k(v,\epsilon) \ge 1-\epsilon_1.$

\begin{figure}[hpbt]
  \centering  
  \epsfig {file=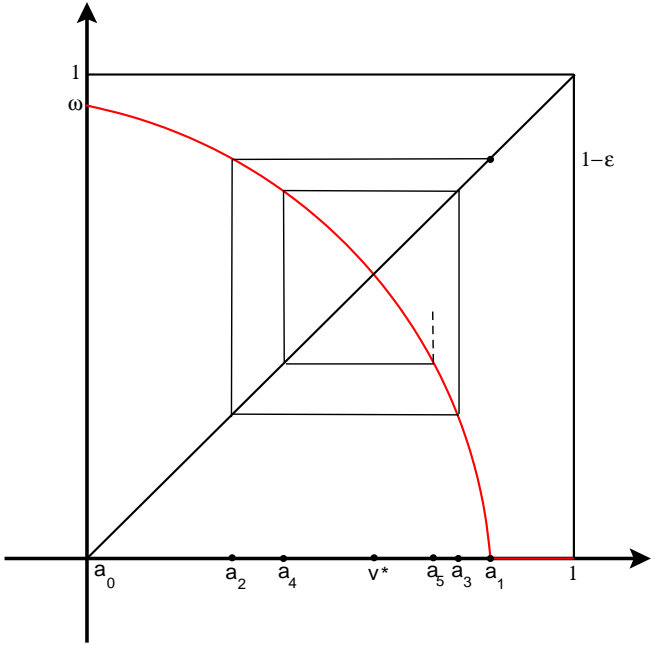, width=3.4in}
  \caption{Boundaries, $a_i,$ of    rate intervals are seen.} \label{Fig2}
\end{figure}

	Next,  we investigate  the rate of synchronization. 	  Set  $a_0 = L_D^{-1}(\omega) =0, a_{k+1} =  L_D^{-1}(a_k), k = 0,1,2,\ldots$ (See Figure \ref{Fig2}). 

 Denote by $S_k$ the region of $[0,1],$  which  points $v$ are synchronized after exactly $k-$ iterations of the map $L_D.$
	One can  see that $S_0 = [1-\epsilon_1,1], S_1 = [a_0,a_2]$  and  $S_k = (a_{k-1}, a_{k+1}],$ if $k \ge 3,$ is an odd  positive  integer, and   $S_k = [a_{k+1}, a_{k-1}),$ if $k \ge 2,$ is an even  positive integer.   One can  see that  $a_k \rightarrow v^*$ as $k \rightarrow \infty.$
	We  shall call $S_k, k \ge 0,$ the rate intervals.

	From the discussion  has been made above it follows that there is no a  finite  time such that { \it all} points of the unite square synchronize. The closer  $v$ is  to  the equilibrium $v^*$ the later is  the moment of synchronization. 
	
		Set $T=\frac{1}{\gamma} \ln \frac{\kappa}{\kappa -1}$ and denote  by $\tilde T$  the time needed for  solution $u(t,0,v^*)$  of  the equation $u'= S - \gamma u,$  to  achieve threshold. Since  all oscillators fire within an  interval of length $T$  and  the distance between two  firing moments of an oscillator  are not   less  than $\tilde T,$  we can  conclude that   the  following  theorem  is  correct.

	\begin{theorem}  Assume that  (\ref{4}) and (\ref{10}) are valid. 
	 If   $t_0\ge 0$  is  a firing moment,   $x_1(t_0) = 1, x_1(t_0+) = 0,$  and  $x_2(t_0+)  \in  S_m$ for some  natural  number $m,$ then the  couple $x_1,x_2$  of  continuously coupled  identical  biological oscillators synchronizes within  the time interval $[t_0 +  \frac{m}{2}\tilde T,t_0 + Tm].$ 
	\end{theorem}  
	\section{Non-identical oscillators: the general case.}

To make  our investigation closer to  the real  world problems one has to  consider an ensemble of non-identical oscillators.  We will  discuss  the following system of equations 
\begin{eqnarray}
 \label{21}
 && 	x'_i = (S +\mu_i)  - (\gamma +\zeta_i)x_i,
 \end{eqnarray}
 where $ 0 \le x_i \le 1 +\xi_i, i = 1,2,\ldots,n.$  The constants $S$ and $\gamma$ are the same as in the last  section such  that  $\kappa = \frac{S}{\gamma} > 1.$ Moreover, constants $\mu_i$ and $\zeta_i$ are sufficiently  small for $\kappa_i = \frac{S+\mu_i}{\gamma+\zeta_i} > 1.$   When $x_j(t) = 1 + \xi_j$ then  the  oscillator fires,   $x_j(t+) = 0.$ 
 The firing changes  values of other  oscillators $x_i, i \not =j,$   such that 
\begin{eqnarray}
 \label{22}
 && x_i(t+) = 0,\, {\rm if}\, x_i(t) \ge  1  - \epsilon
  \end{eqnarray} 
and, if $  x_i(t) < 1  - \epsilon,$ then 
  	 \begin{eqnarray}\label{23}
x_i(t +\tau+)= 
x_i(t +\tau) + \epsilon + \epsilon_i.
\end{eqnarray}
In what  follows, we call  real numbers  $\epsilon, \mu_i,\zeta_i, \xi_i, \epsilon_{i},$ {\it parameters}, assuming  the  first one is positive.  Moreover,  constants $\mu_i,\zeta_i, \xi_i,\epsilon_{i}$  will be called {\it parameters of perturbation}.  Assume that  they  are zeros  to  obtain   the model of {\it identical oscillators}.  We  have that an exhibitory  model  is under discussion, that  is   $\epsilon + \epsilon_{i} > 0$ for all $i.$   Coupling is all-to-all such  that each   firing elicits  jumps in   all non-firing oscillators.   If several oscillators fire simultaneously, then  other oscillators react  as it just  one oscillator fires. In other words, any  firing  acts  only  as a signal  which abruptly  provokes  a state  change, the intensity  of the signal is not important,  and pulse strengths are not additive. 
We have  that 
\begin{eqnarray}
	&& x_i(s) =  x_i(t) {\rm e}^{-(\gamma + \zeta_i)(s-t)} +  \int_{t}^s {\rm e}^{-(\gamma + \zeta_i)(s-u)}(S + \mu_i)du,
	\label{24}
\end{eqnarray}
near $t.$
 
If one  assume that condition (\ref{4}) is valid,  and  constants $\mu_i$ and $\zeta_i$ are sufficiently small such that 
\begin{eqnarray}
	&& \frac{\kappa_i -1}{\kappa_i -1 + \epsilon} < {\rm e}^{-(\gamma + \zeta_i) \tau},
	\label{25}
\end{eqnarray}  
then $x_i(s) <1$ for all $s\in [t,t+\tau],$ if $x(t) < 1 - \epsilon.$   

 In this section we  begin with  analysis of   a couple of oscillators  of the ensemble of $n$  oscillators, and find that the couple synchronizes if    parameters close to  zero.  Then   synchronization of the ensemble will  be proved.

  Consider  the model of $n$ non-identical oscillators   given by  relations (\ref{1})  and (\ref{2}). Fix two  of them, let say,  $x_l, x_r.$ 
    
   \begin{lemma}\label{lem} Assume that conditions  (\ref{4})and (\ref{10}) are valid,
    $t_0\ge 0$  is  a firing moment   such  that   $x_l(t_0) = 1  + \xi_i, x_l(t_0+) = 0.$  If  parameters  are sufficiently close to  zero,  and  absolute values of  parameters of perturbation  are sufficiently small with  respect to  $\epsilon,$   then the couple $x_l, x_r$ synchronizes within the time interval $[t_0, t_0 + T]$ if  $x_r(t_0+) \not \in [a_0,a_{1})$  and  within the time interval  $[t_0 + \frac{m-1}{2}\tilde T, t_0 + (m+1)T],$ if  $x_r(t_0+) \in S_{m},  m \ge 1.$ 
\end{lemma}

{\bf Proof}.   If  $1 + \xi_r-\varepsilon - \varepsilon_{r} \le x_r(t_0)\le  1 +\xi_r,$ then  two oscillators fire simultaneously, and we  have only  to  prove the  persistence of the synchrony,  that will be discussed later. So, fix another  oscillator $x_r(t)$ such  that $ 0 \le  x_r(t_0) < 1 +\xi_r -\varepsilon - \varepsilon_{r}.$
  While the couple  is  not synchronized, there is a sequence of firing moments, $t_i,$  such that  $0 \le t_0<t_2<\ldots,$  and   oscillator $x_l$ fires  at $t_i,$  with  $i$  even,  and $x_r$ fires at $t_i$   with  odd $i.$   For the sake of brevity  let $u_i = x_l(t_i), i = 2j +1,  u_i = x_r(t_i), i = 2j, j \ge 0.$ 
  In what follows we shall  show how to  evaluate $u_{i+1}$ through $L(u_i).$  Consider $i$ is even.
   	 There are $k \le n-2$ distinct firing moments of the motion   $x(t)$  in  the interval $(t_i,t_{i+1}).$ Denote by   $t_i <\theta_1<\theta_2<\ldots<\theta_k <t_{i+1},$   the moments  of firing, when at least one of the coordinates of $x(t)$ fires. We have that 
  \begin{eqnarray} \label{4g3}
&&x_r(\theta_1 +\tau) = (x_r(t_i+\tau) + \epsilon +\epsilon_r){\rm e}^{-(\gamma+\zeta_r)(\theta_1+\tau -t_i)} + \nonumber\\
&&\kappa_r(1-{\rm e}^{-(\gamma+\zeta_r)(\theta_1+\tau -t_i)}), \nonumber\\
&& x_r(\theta_2+\tau) = (x_r(\theta_1+\tau) + \epsilon +\epsilon_r){\rm e}^{-(\gamma+\zeta_r)(\theta_2 -\theta_1)} + \nonumber\\
&&\kappa_r(1-{\rm e}^{-(\gamma+\zeta_r)(\theta_2 -\theta_1)}),\nonumber\\
&&  \ldots\ldots\nonumber\\
&& x_r(\theta_j+\tau) = (x_r(\theta_{j-1}+\tau) + \epsilon +\epsilon_r){\rm e}^{-(\gamma+\zeta_r)(\theta_j -\theta_{j-1})} +\nonumber\\ &&\kappa_r(1-{\rm e}^{-(\gamma+\zeta_r)(\theta_j -\theta_{j-1})}), \nonumber\\
&&  \ldots\ldots\nonumber\\
&& x_r(t_{i+1}) = (x_r(\theta_k+\tau) + \epsilon +\epsilon_r){\rm e}^{-(\gamma+\zeta_r)(t_{i+1} -\theta_k-\tau)} +\nonumber\\
&& \kappa_r(1-{\rm e}^{-(\gamma+\zeta_r)(t_{i+1} -\theta_k-\tau)}).
\end{eqnarray} 
The moment $t_{i+1}$  satisfies 
\begin{eqnarray} \label{4g5}
&& 1 +\xi_r - \epsilon -\epsilon_r  \le x_r(t_{i+1}) \le 1 +\xi_r,
\end{eqnarray}
and  continuously depends on  parameters and $x_r(t_{i}).$ 

We have also that 
\begin{eqnarray} \label{4g4}
&& x_l(\theta_1+\tau) =  \kappa_l(1-{\rm e}^{-(\gamma +\zeta_l)(\theta_1 +\tau-t_i)}), \nonumber\\
&& x_l(\theta_2+\tau) = (x_l(\theta_1+\tau) + \epsilon +\epsilon_l){\rm e}^{-\gamma(\gamma +\zeta_l)(\theta_2 -\theta_1)} +\nonumber\\ &&\kappa_l(1-{\rm e}^{-(\gamma +\zeta_l)(\theta_2 -\theta_1)}), \\
&&  \ldots\ldots\nonumber\\
&& x_l(\theta_j+\tau) = (x_l(\theta_{j-1}+\tau) + \epsilon+\epsilon_l){\rm e}^{-(\gamma +\zeta_l)(\theta_j -\theta_{j-1})} + \nonumber\\
&& \kappa_l(1-{\rm e}^{-(\gamma +\zeta_l)(\theta_j -\theta_{j-1})}), \nonumber\\
&&  \ldots\ldots\nonumber\\
&& x_l(t_{i+1}) = (x_l(\theta_k+\tau) + \epsilon+\epsilon_l){\rm e}^{-(\gamma +\zeta_l)(t_{i+1} -\theta_k-\tau)} + \nonumber\\
&&\kappa_l(1-{\rm e}^{-(\gamma +\zeta_l)(t_{i+1} -\theta_k-\tau)}).\nonumber
\end{eqnarray} 
The last two formulas describe the  dependence  of  $u_{i+1}$ on $u_i.$  One can  easily  find that  similar   dependence  can  be found  if  $i$  is  odd. 

Set $\delta_i(\mu_i,\zeta_i) =  \kappa_i - \kappa.$ One can see that  $\delta_i(0,0) = 0.$ 
Use (\ref{4g3}) and (\ref{4g4}) to  obtain 
\begin{eqnarray} \label{6g3}
&& x_r(t_{i+1}) = (x_r(t_i+\tau) + \epsilon){\rm e}^{-\gamma(t_{i+1} -t_i) } {\rm e}^{-\zeta_r(t_{i+1} -t_i) }+ \nonumber\\
&&\kappa(1-{\rm e}^{-(\gamma + \zeta_r)(t_{i+1} -t_i) }) + 
 \epsilon_r {\rm e}^{-\gamma(t_{i+1} -t_i) } {\rm e}^{-\zeta_r(t_{i+1} -t_i) } +\nonumber\\ 
&&(\epsilon +\epsilon_r)\sum_{j=1}^k {\rm e}^{-(\gamma + \zeta_r)(t_{i+1} -\theta_j-\tau)} +
\delta_r(1 - {\rm e}^{-(\gamma + \zeta_r)(t_{i+1} -t_i) }),
\end{eqnarray} 
and  
\begin{eqnarray} \label{6g4}
&& x_l(t_{i+1}) =  (\kappa + \delta_l)(1-{\rm e}^{-(\gamma + \zeta_l)(t_{i+1} -t_i) })
+ \nonumber\\
&&(\epsilon +\epsilon_l)\sum_{j=1}^k {\rm e}^{-(\gamma + \zeta_l)(t_{i+1} -\theta_j -\tau)}.
\end{eqnarray} 
Now, recall   map $L_D$   defined in the last  section. 
  We have  
 \begin{eqnarray} \label{5g3}
&& \phi(\bar t_{i+1}) = (x_r(t_i +\tau) + \epsilon ){\rm e}^{-\gamma(\bar t_{i+1} - t_i-\tau)} +  \nonumber\\
&&\kappa(1-{\rm e}^{-\gamma(\bar t_{i+1} - t_i)}),
\end{eqnarray} 
where $\bar t_{i+1}$ satisfies
\begin{eqnarray} \label{5g5}
&& \phi(\bar t_{i+1}) =1,
\end{eqnarray}
 and 
\begin{eqnarray} \label{5g4}
&& \psi(\bar t_{i+1}) =  \kappa(1-{\rm e}^{-\gamma(\bar t_{i+1} -t_i)}).
\end{eqnarray} 

By applying  the definition of $L_D$  one can see that   $L_D(u_{i}) = \psi(\bar t_{i+1}).$ 

Assume, without loss of generality,  that  $\bar t_{i+1}  \le t_{i+1}.$ Then one has that 
\begin{eqnarray}\label{hi-hi}
&&   \phi(\bar t_{i+1})-x_r(\bar t_{i+1})  =  1- x_r(\bar t_{i+1})=  \nonumber\\
&&\Phi_1(\epsilon,\epsilon_r,\zeta_r,\delta_r,\tau),
\end{eqnarray}
	where
	\[\Phi_1(\epsilon,\epsilon_r,\zeta_r,\delta_r,\tau)= [\kappa(1-{\rm e}^{-\gamma (\bar t_{i+1} -t_i)})-\]\[(x_r(t_i+\tau) + \epsilon){\rm e}^{-\gamma(\bar t_{i+1} -t_i) }]( {\rm e}^{-\zeta_r(\bar t_{i+1} -t_i) } -1)- 
 \epsilon_r {\rm e}^{-\gamma(\bar t_{i+1} -t_i) } {\rm e}^{-\zeta_r(\bar t_{i+1} -t_i) } - \]\[
(\epsilon +\epsilon_r)\sum_{j=1}^k {\rm e}^{-(\gamma + \zeta_r)(\bar t_{i+1} -\theta_j-\tau)} -
\delta_r(1 - {\rm e}^{-(\gamma + \zeta_r)(\bar t_{i+1} -t_i) }),\]
and the last  expression tends to  zero as  all of its  arguments tend to  zero. 
Next, by  utilizing   (\ref{4g5}) and (\ref{hi-hi})
 we have that  
	$t_{i+1}  - \bar t_{i+1} \le \Phi_2(\epsilon,\epsilon_r,\zeta_r,\delta_r),$ where	
	\[\Phi_2(\epsilon,\epsilon_r,\zeta_r,\delta_r,\tau) \equiv 
	 \frac{ |\xi_r| + \epsilon + |\epsilon_r| + \Phi_1(\epsilon,\epsilon_r,\zeta_r,\delta_r,\tau)}{S-|\mu_r|-\gamma - |\zeta_r|}.\]
 Now,  by  applying  the last  inequality, (\ref{6g4}) and (\ref{5g4})  one can   see that   \[| L_D(u_{i})-K_i(u_{i})| =
  |x_l(t_{i+1}) - \psi(\bar t_{i+1})|\le |x_l(t_{i+1}) - x_l(\bar t_{i+1})| + \]\[|x_l(\bar t_{i+1}) - \psi(\bar t_{i+1})| \le \Phi_2(S+|\mu_l|+\gamma  +|\zeta_l|) + \Phi_1.\] 
 That  is,  difference $L_D(u_{i},\epsilon)- u_{i+1}$  can be made  arbitrarily  small if the parameters are sufficiently close to zero. Moreover, we should assume smallness of  absolute values of the parameters of perturbation with  respect to  $\epsilon,$ to  satisfy (\ref{4g5}). 
This convergence is uniform  with  respect to  $u_0.$ We can also  vary the number of points $\theta_i$ and their location  in the intervals $(t_j,t_{j+1})$ between $0$ and $n-1.$ The  convergence is indifferent with  respect to  these variations, too. 
 
  Consider $L_D^i(u_0,\epsilon).$ It is true that  $L_D^m(u_0,\epsilon) \in [1-\epsilon,1].$ Assume, without lost  of generality, that $m$ is an even number. Since $L_D$ is a continuous function, we can  find  recurrently, by  applying    
  the following  sequence of inequalities
$|u_i - L_D^i(u_0,\epsilon)| \le |u_i - L_D(u_{i-1},\epsilon)| + | L_D(u_{i-1},\epsilon) -L_D(L_D^{i-1}(u_0,\epsilon))|, i = 1,2,\ldots,$   
  that  either $1+\xi_r-\epsilon-\epsilon_r < u_m < 1+\xi_r$ or  $1+\xi_l-\epsilon-\epsilon_l < u_{m+1} < 1+\xi_l,$ if the parameters are sufficiently  small.   From the notation it implies that  each  of the  last   two  inequalities bring the  couple to  synchronization. 
 
     Since each of  the iterations of  $L_D$  is done  within interval with length not more than $T,$ we obtain now that the couple $x_l,x_r$ is synchronized not later than  $t= t_0+ (m+1)T.$   
   
 We have found  that   oscillators $x_l$ and $x_r$  fire  in unison at  some moment $t = \theta.$  Next, we  show that   they will save the state, being different.  To find conditions for this, let  us  denote by   $\tau > \theta$  the next  moment of firing of the couple. Let say, $x_r$ fires at  this moment.  Thus, we have that $x_l(\theta+)  =x_r(\theta+) =0.$  Then  $x_l(t)  =x_r(t), \theta \le t \le \tau.$  It is clear that to  satisfy $x_l(\tau+)  =x_r(\tau+) =0,$ we need 
$1 + \xi_r - \epsilon - \epsilon_r \le x_l(\tau).$ By applying  formula  (\ref{4g5}) again, this time with  $t_i = \theta, t_{i+1} = \tau,$ one can  easily  obtain that  the inequality  is correct if  parameters   are close to  zero and  absolute values of the parameters of perturbation  are small with  respect to  $\epsilon.$   Thus, one can  conclude that  if  a couple of oscillators is synchronized at   some moment of time than it continues  to  fire in unison for ever. The lemma is proved.

 Let us  extend the result of the last  Lemma  for the whole ensemble.  
 \begin{theorem}\label{thm}   Assume that  (\ref{4})  and (\ref{10}) are valid
, $t_0\ge 0$  is  a firing moment   such  that   $x_j(t_0) = 1  + \xi_j, x_j(t_0+) = 0.$ If the  parameters  are sufficiently close to  zero,  and  absolute values of  parameters of perturbation  are sufficiently small with  respect to  $\epsilon,$  then   the motion   $x(t)$ of the system  synchronizes  within the time  interval  $[t_0,t_0 +T],$ 
if   $ x_i(t_0+) \not \in [a_0,a_1), i \ne j,$   and  within the time interval  $[t_0 + \frac{\max_{i \ne j}k_i-1}{2}\tilde T,t_0 +(\max_{i \ne j}k_i+1)T],$  if  there exist  $x_s(t_0+) \in [a_0,a_1)$ for some $s \not = j$  and  $x_i(t_0+) \in S_{k_i}, i \not = j.$
\end{theorem} 

{\bf Proof.} Consider the  collection of couples $(x_i,x_j), i \ne j.$   Each  of these pairs synchronizes by the last  Lemma  within  interval  $[t_0 + \frac{\max_{i \ne j}k_i-1}{2}\tilde T,t_0 +(\max_{i \ne j}k_i+1)T].$ The  theorem is proved. 

Let us  introduce  a more general system of  oscillators  such  that  Theorem \ref{thm} is still true. 

Consider  a system of $n$ oscillators  given such that  if $i-$th oscillator does not fire or jump up, it satisfies $i-$th equation  of system (\ref{1}). If  several oscillators $x_{i_s}, s = 1,2,\ldots,k,$ fire  such  that $x_{i_s}(t) = 1 + \phi(t,x(t),x(t - \tau_{i_s}),$ where $|\phi(t,x(t),x(t - \tau_{i})| <  \xi_{i}, i = 1,2,\ldots,n,$ and  $x_{i_s}(t+) = 0,$ then all other oscillators $x_{i_p}, p= k+1,k+1,\ldots,n,$ change their coordinates by  law
\begin{eqnarray}
 \label{225}
 && x_i(t+) = 0,\, {\rm if}\, x_i(t) \ge  1  - \epsilon
  \end{eqnarray} 
and, if $  x_i(t) < 1  - \epsilon,$ then 
  	 \begin{eqnarray}\label{235}
x_i(t +\tau+)= 
x_i(t +\tau) + \epsilon +  \sum_{s=1}^k \epsilon_{i_pi_s}.
\end{eqnarray}

One can  easily  see that the last  theorem is correct  for the model  just  have been  described, if 
$\epsilon + \sum_{s=1}^k \epsilon_{i_pi_s} > 0,$ for all possible $k,i_p$ and $i_s,$ and we assume that  $ \epsilon_{ij}$ are also  parameters of perturbation.  Moreover,  one can  easily  see that  initial functions  for thresholds conditions  can  be chosen arbitrarily  with  values in the domain of the system. 

\begin{remark} 
  Our preliminary  analysis  shows  that  the dynamics  in  a neighborhood of $v^*$  can be very  complex. We do not exclude  that a chaos  appearance can be observed, and trajectories   may   belong to  a fractal,  if parameters are not small.  It does not contradict the  zero  Lebesgue measure of non-synchronized  points.  Possibly, analysis  of non-identical oscillators  with not small  parameters is of significant interests to  explore arrhythmias, earthquakes,  chaotic flashing of fireflies, etc. 
\end{remark}
 
\begin{remark}  The time of synchronization  for a given initial point does not increase if number  of oscillators increases (but the parameters needed to be closer to zero). This property, possibly,   can  be accepted as  a small-world phenomenon.
\end{remark}
\section{The simulation result}

\begin{figure}[hpbt]
  \centering  
  \epsfig {file=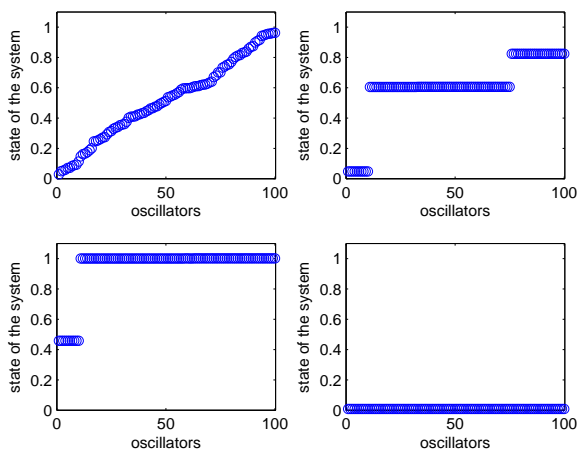, width=3.4in}
  \caption{  The  state of the model   before the first, the $183-$th, $366-$th, and     $549-$th jump  is seen. The flat  fragments  of the graph are groups of  oscillators firing in unison. } \label{Fig3}
\end{figure}

To  demonstrate our main result  numerically,  let  us consider a model of $100$ oscillators,  which initial  values are randomly  uniform  distributed  in  $[0,1].$   Their  differential  equations are of form 
$$x_i' =  ( 4.1+0.01*sort(rand(1,n))  - ( 3.2+0.01*sort(rand(1,n))x_i,$$  and thresholds  $$1+0.005*sort(rand(1,n)), i = 1,2,\ldots,100,$$ where deviations  of coefficients  the threshold  are also  uniformly  random   in  $[0,1].$  We place the result of simulation  with  $\epsilon = 0.06$ and $\tau = 0.002$ in Figure \ref{Fig3},  where the  state of the system is shown  at the initial moment,  before  the $183-$th jump,  before the $366-$th jump and the last is before the $549-$th jump.  That  is, it is obvious that  eventually all oscillators fire  in unison.       
	
 \section{Conclusion} The cardiac pacemaker model of  identical and non-identical oscillators  with  delayed pulse-couplings is investigated in  the paper.  We apply the method of investigation proposed in \cite{akhmet1}, which  is based on a specially  defined map. The map  is, in fact  a Poincar\'e map  if  one considers the identity  of oscillators.   Sufficient conditions are found   such  that  delay  involvement in the Peskin's model  does not  change the  synchronization result for  identical  and non-identical  oscillators \cite{p,ms,akhmet1}.   The result has a biological  sense, since retardation is often  presents in biological  processes and  if one proves that   a phenomenon preserves even with  delays, that  makes us more  confident   that  the model  is adequate to the reality.   Moreover, the method of treatment of  models with  delay   can  be useful  for  neural  networks and  earthquake faults \cite{epg,gerst, herz, hopfield, olami} analysis.  All proved  assertions can  be easily  specified  with  $\tau = 0,$ to  obtain synchronization of the Peskin's model for identical \cite{p,ms} and nonidentical  \cite{akhmet1} oscillators. In particular,  from the theorem  of   Section \ref{sec2} one can obtain  the result  of Example $2.2$ of   \cite{akhmet1}.

\end{document}